\documentclass{ws-procs9x6}

\begin{document}

\title{Measurement of the $Q^2$-evolution of the Bjorken integral
at low $Q^2$}

\author{A. Deur}

\address{Thomas Jefferson National Accelerator Facility \\
12000 Jefferson Avenue \\ 
Newport News, VA 23606, USA\\ 
E-mail: deurpam@jlab.org}

\maketitle

\abstracts{We report on the extraction of the $Q^2$-dependence 
of the Bjorken sum between $0.16<Q^2<1.1$ GeV$^2$. 
A twist analysis performed on these data shows that
the higher twist corrections are small due to a 
cancellation between the twist-4 and 6 terms. The extraction
of an effective strong coupling constant is discussed.
}

\vspace{-1cm}
\section{Bjorken Sum Rule}

The Bjorken sum rule\cite{Bjorken} played a central role in 
verifying in the spin sector that QCD is the correct gauge theory of
strong interaction. It reads:
\begin{equation}
\int_{0}^{1}(g_{1}^{p}-g_{1}^{n})dx=\frac{g_{a}}{6}
[ 1-\frac{\alpha_{\rm{s}}}{\pi}-3.58\left(\frac{\alpha_{\rm{s}}}
{\pi}\right)^{2}-20.21\left(\frac{\alpha_{\rm{s}}}{\pi}\right)^{3} +...]
+\sum_{i=1}^\infty {\mu^{p-n}_{2i+2} \over Q^{2i}}
\label{eqn:bj}
\end{equation}
\noindent 
where the coefficients $\mu^{p-n}_{2i+2}$ are sums of elements of twist no
higher than $2i$ and which $Q^2$-dependence is given by DGLAP equations. 
The sum rule has been verified experimentally at $Q^2$=5 GeV$^2$ to better 
than 10\%.

Taking the $Q^2 \rightarrow 0$ limit relates the Bjorken sum rule to the 
Gerasimov-Drell-Hearn (GDH) sum rule\cite{GDH} that stands at $Q^2=0$. 
This connection triggered the generalization of the GDH sum\cite{gGDH} 
at finite $Q^2$. 
Because the generalized GDH sum is in principle calculable 
at any $Q^2$, it can help to study the transition from
the hadronic to partonic degrees of freedom of strong interaction.  
However, the validity domains of the chiral perturbation theory ($\chi$PT) 
at low $Q^2$ and pQCD calculations at higher $Q^2$ do not overlap. 
Lattice QCD should bridge  
the two domains but no calculation is available yet. The relation 
between the Bjorken and generalized GDH sum is:
\begin{equation}
\int_{0}^{1^{-}}g_{1}^{p}-g_{1}^{n}dx=
\frac{Q^2}{16\pi^2\alpha}(\rm{GDH}^p(Q^2)-\rm{GDH}^n(Q^2))
\label{eqn:link}
\end{equation}
\noindent 
Hence the Bjorken sum is essentially the $p-n$ flavor non-singlet part of the GDH sum,
which yields simplifications: more reliable estimations of the unmeasured low-$x$ part 
of the  integral, simpler pQCD evolution equations and less complicated
$\chi$PT calculations. This might help in linking validity domains  of pQCD and 
$\chi$PT\cite{Volker}. Hence the Bjorken sum appears as an important quantity 
to measure to understand the hadron-parton transition. 

Precise data on the proton\cite{eg1a proton}, deuteron\cite{eg1a deuteron}, 
and $^3$He\cite{E94010} are available from the Thomas Jefferson National
Accelerator Facility (JLab). We used them to extract the Bjorken sum from 
$Q^2=0.16$ to 1.1 GeV$^2$. To combine proton and neutron 
data, the $^3$He data were reanalyzed at the same $Q^2$ as the proton data. 
For consistency, the unmeasured low-$x$ part of 
the integral was re-evaluated for the three data sets using a consistent 
prescription. The results are shown on Fig. 1, left panel, 
together with SLAC E143 results in the resonance region\cite{E143}. 
The elastic contribution is not included. The systematic uncertainties 
are given by the horizontal bands. At low $Q^2$ $\chi$PT calculations can be 
compared to the data. 
Calculation done in the heavy Baryon approximation\cite{jichipt} may be more 
reliable since they agree with data, up to 
about $Q^2=0.25$ GeV$^2$ (to be compared to $Q^2=0.1$ GeV$^2$  typically for singlet
quantities). However, the results from Bernard \emph{et al.}\cite{meissnerchipt}
which do not use this approximation, do not support this conclusion. 
The improved model of Soffer and 
Teryaev\cite{ST} and the calculation from Burkert and Ioffe\cite{AO} agree well with 
the data. The pQCD result at third order in $\alpha_s$ and leading twist is shown 
by the 
gray band. The width comes from the uncertainty on $\alpha_s$. 
Somewhat surprisingly, the data agree with the leading twist calculation down to 
quite 
low $Q^2$, which indicates that higher twist terms are small or cancel each 
others. We quantitatively addressed the question by performing a
higher twist analysis\cite{bj} that shows that the coefficient $\mu^{p-n}_4$ 
of the $1/Q^2$ correction has opposite sign and similar magnitude to the 
coefficient $\mu^{p-n}_6$ of the $1/Q^4$ correction at $Q^2=1.$ GeV$^2$:
$\mu^{p-n}_6/Q^4=0.09 \pm 0.02$ and $\mu^{p-n}_4/Q^2 \simeq -0.06 
\pm 0.02$. Higher Twist analyzes were also done separately on the 
proton\cite{HT_P} and neutron\cite{HT_N}. 
\vspace{-0.2cm}
\section{The Effective Strong Coupling Constant}
The simple $\alpha_{\rm{s}}$-dependence of the Bjorken sum
makes it an ideal tool to extract $\alpha_{\rm{s}}$. However, higher
twists have to be known or negligible. 
This is not the case for us. This difficulty is avoided when using the concept of
effective coupling constants\cite{grunberg}. Here, higher twists and
QCD radiations at orders higher than $\alpha_{\rm{s}}^2$ are folded into
$\alpha_{\rm{s}}^{\rm{eff}}$. We obtain a quantity that is
well defined at any $Q^2$, well behaved when crossing $\Lambda_{\rm{QCD}}$ or
a quark mass threshold and that is renormalization scheme independent. However,
$\alpha_{\rm{s}}^{\rm{eff}}$ becomes process dependent. This is not a problem  
since process-dependent coupling constants can be related using  
equations called ``commensurate scale relations'' that connect observables 
without scheme or scale ambiguity\cite{brodsky1,brodsky11}. The extracted 
$\alpha_{\rm{s}}^{\rm{eff}}$ is shown
 in Figure~1, right panel. Also plotted is the pQCD calculation of 
$\alpha_{\rm{s}}^{\rm{eff}}$ (light gray band), $\alpha_{\rm{s}}$ calculated to
order $\beta_0$ (pink band), $\alpha_{\rm{s}}^{\rm{eff}}$ calculated using the
model of Burkert and Ioffe, and $\alpha_{\rm{s}}^{\rm{eff}}$ as extracted from 
SLAC E155 (open square). Satisfactorily, $\alpha_{\rm{s}}^{\rm{eff}}$ merges with 
$\alpha_{\rm{s}}$ at large $Q^2$ as expected since their difference
is due to pQCD radiative corrections and higher
twists. At low $Q^2$, the behavior of $\alpha_{\rm{s}}^{\rm{eff}}$ is
constrained by the GDH sum rule. This, together with the 
steep rise $\alpha_{\rm{s}}^{\rm{eff}}$ at larger $Q^2$ strongly hint that
$\alpha_{\rm{s}}^{\rm{eff}}$
has no significant scale dependence at low $Q^2$. This possible ``freezing'' of 
$\alpha_{\rm{s}}^{\rm{eff}}$ at low $Q^2$ is a debated issue. Lower $Q^2$ 
data\cite{E97110,E03006}will have the definite word on this feature of 
$\alpha_{\rm{s}}^{\rm{eff}}$.   
\vspace{-0.5cm}
\begin{figure}[ht]
\centerline{\epsfxsize=4.5in\epsfbox{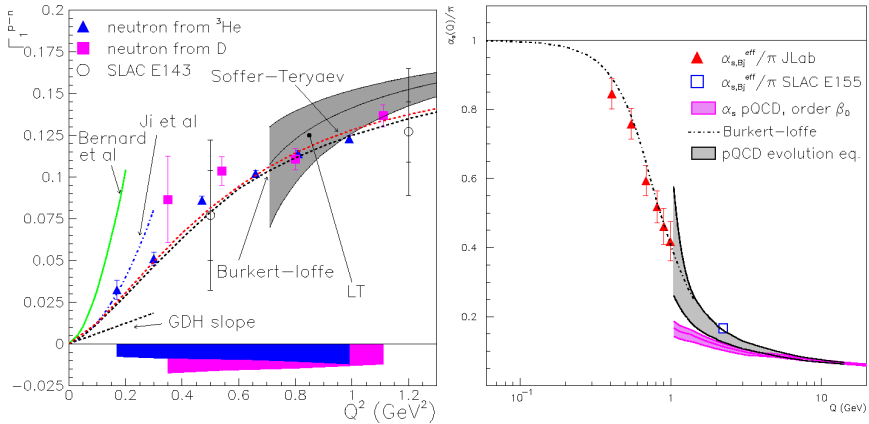}}   
\caption{Left: $Q^2$-evolution of the Bjorken sum. The Bjorken sum formed
using neutron data extracted from $^3$He (D) data is shown by the triangles 
(squares). Right: Effective strong coupling constant.  
\label{bj}}
\end{figure}
\vspace{-1cm}
\section{Summary and Outlook}
We have extracted the Bjorken sum in the $Q^2$ range of 0.16-1.1 GeV$^2$. The 
parton to hadron gap, if smaller, is not bridged yet.
The magnitudes of the higher twists were extracted. The 
higher twist effects appear to be small, due to a cancellation of the
1/Q$^2$ and 1/$Q^4$ terms. We extracted a physical coupling for the 
strong interaction. Data hint that $\alpha_{\rm{s}}^{\rm{eff}}$ loses it scale 
dependence at low $Q^2$. 
\vspace{-0.2cm}
\section*{Acknowledgments}
This work was supported by the U.S. Department of Energy (DOE) and the U.S.
National Science Foundation. The Southeastern Universities Research 
Association operates the Thomas Jefferson National Accelerator 
Facility for the DOE under contract DE-AC05-84ER40150.

\end{document}